\documentclass[]{spie}  %>>> use for US letter paper
\usepackage{amsmath,amsfonts,amssymb}
\usepackage{graphicx}
\usepackage{multirow}
\usepackage[colorlinks=true, allcolors=blue]{hyperref}

\newcommand{\ff}[1] {  \mbox{\footnotesize{#1}}  }

\title{Accurately identifying vertebral levels in large datasets} 

\author[a]{Daniel C. Elton}
\author[a]{Veit Sandfort}
\author[b]{Perry J. Pickhardt}
\author[a]{Ronald M. Summers}
\affil[a]{Imaging Biomarkers and Computer-Aided Diagnosis Laboratory\\
Radiology and Imaging Sciences\\
National Institutes of Health Clinical Center\\
Bethesda, MD 20892-1182, USA}
\affil[b]{School of Medicine and Public Health\\
University of Wisconsin, Madison, USA}
\authorinfo{Further author information: Send correspondence to Daniel Elton (\href{mailto:daniel.elton@nih.gov}{daniel.elton@nih.gov}) or Ronald Summers \href{mailto:rms@nih.gov}{rms@nih.gov}}

\begin{document} 
\maketitle

\begin{abstract}
%max 250 words 
The vertebral levels of the spine provide a useful coordinate system when making measurements of plaque, muscle, fat, and bone mineral density. Correctly classifying vertebral levels with high accuracy is challenging due to the similar appearance of each vertebra, the curvature of the spine, and the possibility of anomalies such as fractured vertebrae, implants, lumbarization of the sacrum, and sacralization of L5. The goal of this work is to develop a system that can accurately and robustly identify the L1 level in large heterogeneous datasets. The first approach we study is using a 3D U-Net to segment the L1 vertebra directly using the entire scan volume to provide context. We also tested models for two class segmentation of L1 and T12 and a three class segmentation of L1, T12 and the rib attached to T12. By increasing the number of training examples to 249 scans using pseudo-segmentations from an in-house segmentation tool we were able to achieve 98\% accuracy with respect to identifying the L1 vertebra, with an average error of 4.5 mm in the craniocaudal level. We next developed an algorithm which performs iterative instance segmentation and classification of the entire spine with a 3D U-Net. We found the instance based approach was able to yield better segmentations of nearly the entire spine, but had lower classification accuracy for L1.\end{abstract}

%100 word abstract for printed programs
%The vertebral levels of the spine provide a useful coordinate system when making measurements of plaque, muscle, fat, and bone mineral density. Correctly classifying vertebral levels with high accuracy is challenging due to the similar appearance of each vertebra, curvature of the spine, and the possibility of  anomalies such as fractured vertebrae, implants, lumbarization of the sacrum, and sacralization of L5. The goal of this work is to develop a system that can accurately and robustly identify the L1 level in large heterogeneous datasets. The first approach we study is using a 3D U-Net to segment the L1 vertebra directly using the entire scan volume to provide context. We also tested a two class segmentation of L1 and T12 and a three class segmentation of L1, T12 and the rib attached to T12. By increasing the number of training examples to 249 scans using pseudo-segmentations from an in-house segmentation tool we were able to achieve 98\% accuracy with respect to identifying the L1 vertebra with an average error of 4.5 mm in the craniocaudal level. We next developed an algorithm which performs iterative instance segmentation and classification of the entire spine with a 3D U-Net. We found the instance based approach was able to yield better segmentations of nearly the entire spine, but with lower classification accuracy for L1

% Include a list of keywords after the abstract 

%----------------------------------------------------------------------------- 
%\section{SUMMARY}
%100 words
%We compare several different methods for identifying vertebral levels. The first method we studied is using a 3D U-Net to do a segmentation of the transitional vertebra L1. We obtained an accuracy of 88\% and an average level error of 8 mm by training on pseudo segmentations generated by a previously developed code which does segmentation using the watershed algorithm and other techniques. We next implemented a system using a 3D U-Net that performs iterative instance segmentation and classification of the entire spine and compared its performance to the direct approach. 

%-------------------------------------------------------------------------------- 
%\section{PURPOSE}
%The purpose of this work is to develop a deep learning system to accurately identify the L1 level which can operate on large heterogeneous datasets where spinal abnormalities may be present. Such a system would be useful for making consistent automated measurements of bone mineral density, fat, plaque, and other metrics. 
%-------------------------------------------------------------------------------- 
\section{INTRODUCTION}
The application of deep learning systems to radiology holds much promise for making medicine more  quantitative.\cite{Burns2019JBMR,Sahiner2018MedPhys} Deep learning based systems can be used opportunistically to gather valuable information from CT scans which is useful for screening for certain diseases and identifying risk factors. Examples of automated opportunistic measurements that have been demonstrated recently are bone mineral density,\cite{Jang2019Radiology20000,Pickhardt2019population} visceral fat,\cite{Lee2018Fully} muscle,\cite{Graffy2019BJR} organ volume,\cite{Roth2018MIA} plaque burden,\cite{Graffy2019Automated} and body composition analysis.\cite{Weston2019Automated} If a patient has multiple CT scans performed, changes over time can be tracked. The resulting data is not only useful for patient care but also for scientific studies looking at health outcomes in large cohorts. However, in order for such systems to be useful consistent measurement is important. Being a rigid framework, the vertebrae of the spine provide a useful coordinate system for making measurements. Conventions have been established in this regard - for instance it is common to measure muscle content at L3\cite{Graffy2019BJR}, and bone mineral density at L1.\cite{Pickhardt2019population} Aortic plaque burden may be measured between the L1 and L4 levels.\cite{Graffy2019Automated}

The thoracic vertebrae (T1-T12) and lumbar vertebrae (L1-L5) all have a similar appearance so classification based solely on a cropped bounding box around a given vertebra is difficult. Radiologists identify vertebrae based on their position relative to transitional vertebrae such as L1/T12 and L5, and we hypothesized that a deep learning system can learn to do something similar. However, classification and segmentation of the vertebrae is made challenging by the high variability of the spine. One source of variability is varying degrees of curvature due scoliosis/kyphosis or due to the patient not lying perfectly straight in the scanner. Another source of variability is the presence of lumbosacral transitional vertebrae, which occur in about 10\% of patients.\cite{French2014Lumbosacral} About 6\% of patients have lumbarization of S1, and about 4\% of patients have sacralization of L5.\cite{French2014Lumbosacral} For this reason, L5 is not a reliable transitional vertebra to use as a reference landmark. Abnormalities also exist at the T12 and L1 vertebrae, such as the presence of 13th rib bearing vertebral bodies (incidence about 1\%\cite{Aly2015ribs}) or rudimentary/non-existent 12th ribs (incidence about 8\%\cite{Glass2002ribs}). In this work we choose to focus on using L1 as our reference vertebrae. 

Several machine learning based approaches to segmenting and partitioning the spine have already been published. A full review is outside the scope of this paper, but we summarize the best-performing systems we found in table \ref{tab:spinesegpapers}. We noticed that some prior studies (especially pre-2016) used small test sets (ie.\ 2-10 cases) which have a low probability of having abnormalities. One of the state of the art methods is a system developed by Lessmann, et al.\ which uses iterative instance segmentation with a 3D U-Net.\cite{Lessmann2019,Lessmann2018} Overall they achieved a 94\% accuracy (correct identification rate) for classifying lumbar vertebrae. In this paper we compare direct segmentation of L1 with an iterative instance based approach which is similar to the approach developed by Lessmann et al. 

%Within a subset of 5 lumbar spine CTs they had a 100\% accuracy. 
%They note that one patient had only 4 lumbar vertebra, so the labeling in that case was off by one. 
%Cai, M. Landis, D.T. Laidley, A. Kornecki, A. Lum, S. Li Comput. Medical Imag. Graph., 51 (2016), pp. 11-19
%Suzani et al.\ trained a neural network to detect and localize vertebrae by regressing their centroids. While they detection rates of ~95\% their system does not label the vertebrae.\cite{Suzani2015} 

\begin{table}
    \centering
    \label{tab:spinesegpapers}
    \begin{tabular}{p{1.5cm} p{6cm} c c c c}
architecture & details & $N_{\ff{train}}$ & $N_{\ff{test}}$ & acc. &  citation(s) \\ 
\hline
{\footnotesize 3D U-Net }& {\scriptsize	 Trained on spinal CT (33\%) and low dose chest CT (66\%). Patch based iterative instance segmentation.} & 60 & 30 & 94 \% & Lessmann et al. 2019 \cite{Lessmann2019} \\
\hline
{\footnotesize Two stage CNN. }& {\scriptsize Trained on spinal CT. Two stage network consisting of a 3D U-Net for detection of the spine and a 2D U-Net for vertebral identification. }  & 242 & 60 & 92 \% & McCouat \& Glocker 2019 \cite{mccouat2019vertebrae} \\
\hline
{\footnotesize 3D CNN + FCNN + biRNN} & {\scriptsize Trained on spinal CT. Patch based 3D CNN used for initial labeling. Labels are refined using a FCNN coupled to a bidirectional RNN.} & 242 & 60 & 92\% & Liao et al., 2018 \cite{Liao2018IEEETMI} \\
\hline
{\footnotesize 3D U-Net} & {\scriptsize	 Trained on various CT scans. Network performs centroid placement and classification/labeling only. Uses ``deep supervision'', message passing, and sparsity regularization.} & 1112 & 112 & 84\% & Yang et al., 2017 \cite{Yang2017Automatic} \\
\hline
{\footnotesize FCNN + 3D U-Net} & {\scriptsize Trained on spinal CT. A FCNN is used to regress a bound box around the lumbar region, and a multiclass 3D U-Net is used to segment the 5 lumbar vertebrae.} & 12 & 3 & * & Janssens et al.\, 2016 \cite{Janssens2018Fully} \\
\hline
    \end{tabular}
    \caption{Select prior works on vertebrae segmentation and identification using deep learning which report the highest accuracy. $^*$Janssens et al.\ do not perform classification, but their segmentations of the 5 lumbar vertebrae had average Dice coefficients of 0.95 so we can surmise the correct vertebrae could be labeled nearly 100\% of the time, at least within their small test set of 3 scans.}
\end{table}

%-------------------------------------------------------------------------------- 
\section{METHODS}
%--------------------------------------------- 
\subsection{Dataset preparation}
%\cite{Glocker2012AutomaticLocalization,Glocker2013Vertebrae}
To construct a training set of L1 segmentations we initially hand segmented 52 scans using 3D Slicer. 37 of the scans were taken from the ``CT Lymph Nodes'' dataset\cite{CIAlymphnodesdataset,Roth2014} of the Cancer Imaging Archive\cite{Clark2013CIA} and 15 come from a set of scans taken for CT colonoscopy at National Naval Medical Center.\cite{Pickhardt2003CTCData} To generate a larger training data set we used a previously developed automated bone mineral density (BMD) code,\cite{JianhuaYao2006} to generate rough segmentations of the entire spine (``pseudomasks''). To partition the spine, the tool detects the spinal cord, performs a curved reformulation along the spinal cord centerline, and partitions the spine based on the bone density along the centerline. For segmentation the watershed algorithm is used, combined with directed graph search to reduce oversegmentation. The L1 level was identified manually in each case, and the segmentation at the L1 level was extracted from the pseudo-segmentations generated by the code. Manual levels were found for 225 cases from the aforementioned Naval dataset and 32 cases from the Cancer Imaging Archive, to generate L1 segmentations for a total of 257 cases. 

To train the 3D U-Net for iterative instance segmentation the cervical, thoracic, and lumbar vertebra were labeled with integers between 1 and 24 and the labels were aligned so that the vertebra with label 20 matched the level that was determined manually to be L1. To generate the subvolumes, we first resampled all images to 1mm x 1mm x 1mm and clip the images to Hounsfeld units between -100 and 2000. Training and inference are performed with patches of size 128x128x128, which is large enough to include an entire vertebra as well as at least one vertebra above and below. The training subvolumes are generated by centering a 128x128x128 box on each vertebra in the ground truth data, with minor random jitter of $\pm$ 2 cm in each direction. Data augmentation similar to the prior method was used (random XY flipping, random B-spline deformation, and random rotations). The center (target) vertebra is given the label `1' and all vertebra above and below are all given the labels `2' and `3' for use as the instance memory during training. The network can be trained with only the above vertebra in the instance memory (``top-down'' mode), only the ones below (``bottom-up'' mode), or by randomly selecting one or the other (a ``bidrectional'' network). 15-20\% of the generated subvolumes were empty, which is essential in order to train the algorithm to ignore things that would otherwise be mistaken as vertebra, such as contrast or hip bone.

We first trained the instance-based network on the generated pseudomasks of the entire spine for the same 257 cases discussed above. We also trained with 2 scans from the 2019 MICCAI ``VerSe'' challenge, which consist of a wide variety of volumes with hand segmentations of all the vertebra present in the scan.\cite{Sekuboyina2018} Since we are only interested in the lumbar and thoracic vertebrae we disregarded any cervical vertebra in the VerSe data. We also used 10 scans with manual segmentations of the lumbar vertebrae made available by Prof.\ Vrtovec at the Laboratory of Imaging Technologies at the University of Ljubljana.\cite{Ibragimov2014lumbardataset,Korez2015lumbardataset} 
%Another dataset which may be used for increasing diversity is the MICCAI 2016 xVertSeg dataset which contains 16 spinal CT with manual segmentations, many of which contain fractures.\cite{xVertSeg}

To test the accuracy of both approaches for predicting the L1 level we used 40 cases from a separate set of scans taken for CT colonography, where the L1 level was identified manually. 
%Additionally as hold-out test sets for examining the segmentation performance of iterative instance segmentation we used 8 scans from the Naval data and 5 scans from the VerSe data. 

%--------------------------------------------- 
\subsection{Identifying the L1 level}
\begin{figure}[ht]
    \begin{center}
    \includegraphics[height=6cm]{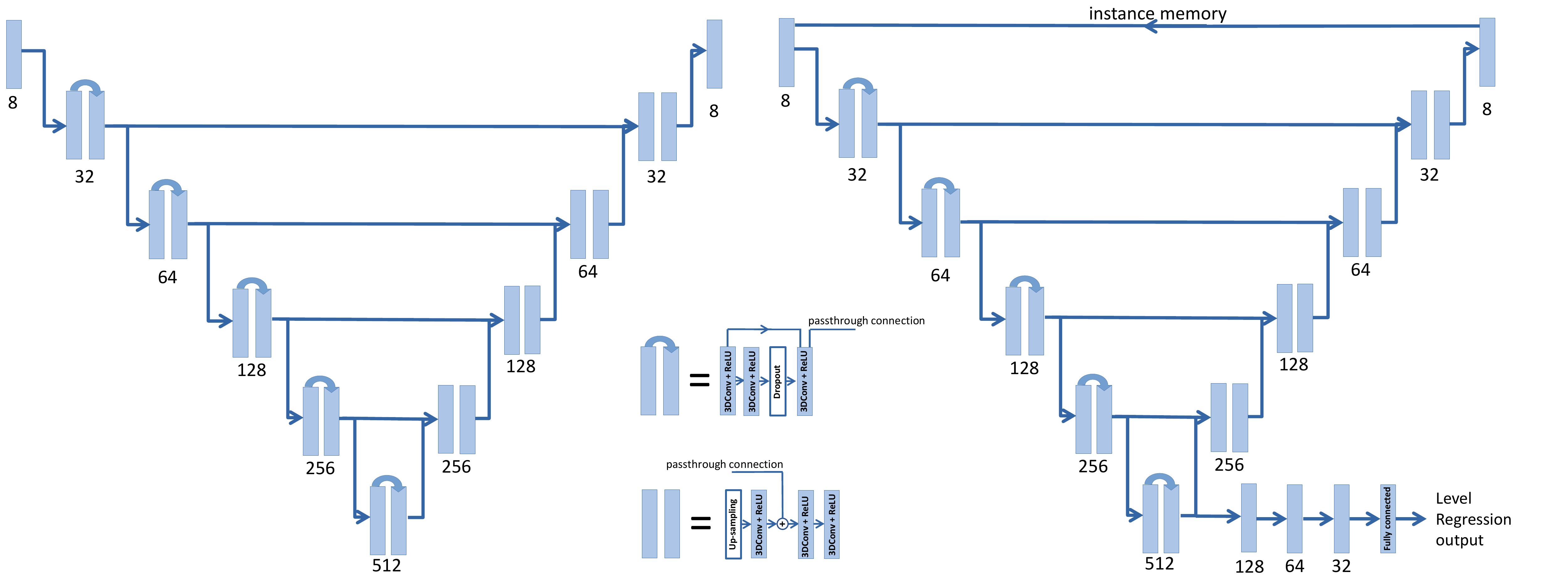}
    \end{center}
    \caption[example] 
    %>>>> use \label inside caption to get Fig. number with \ref{}
    {\label{fig:modelarchs} The 3D U-Net architectures used for identifying L1 (left) and for instance segmentation (right). }
\end{figure} 

We tested segmenting the L1 vertebrae using a 3D U-Net\cite{Ronneberger2015,cicek20163DUnet} architecture (shown in fig.\ \ref{fig:modelarchs}). The 3D U-Net is adapted from an architecture developed for organ segmentation\cite{Sandfort2019Data} and is implemented in Pytorch. Each layer uses group normalization\cite{Wu2018GroupNorm} (group size = 16), dropout (dropout rate = 0.3), and a leaky ReLU activation function. Images are re-sampled to 128x128x192 and clipped to between 100-1000 HU before normalization. For augmentation we used independent random flips along the x, y, and z directions, elastic deformations using the B-spline deformation function in SITK and random rotations between -20 and 20 degrees around a random axis. Training was performed on a 4 Titan X GPUs with a batch size of 4 and run for $\approx$ 100,000 iterations. 

%---------------------------------------------- 
\subsection{Iterative instance segmentation}
We hypothesized that performing a full spine segmentation would result in more robust identification of L1. The approach we took is a simplified and modified version of the iterative instance algorithm developed by Lessmann et al.\cite{Lessmann2019, Lessmann2018} To simplify their approach we do not include a third output branch to detect whether a vertebra is fully or partially visible, and we do not use a loss function which weights the surface of vertebra more heavily. We used our own custom 3D U-Net architecture which is shown in fig.\ \ref{fig:modelarchs}. 
%We tested two different architectures - a modification of hte  architecture similar to the one used for L1 segmentation (shown in figure 1) and the 3D U-Net architecture developed by Isensee et al.\cite{Isensee20183DUnet} for brain tumor segmentation which features ``deep supervision'' on the uspampling side.
% We found this architecture performed better than the architecture used for segmenting L1.

The loss function we used is a weighted sum of the number of false positive voxels ($\mbox{FP}$) and number of false negative voxels ($\mbox{FN}$) plus a term for level regression error: 
\begin{equation}\label{eqn:lossfun}
    \mathcal{L} = \lambda_n\mbox{FP}_{\ff{frac}} + \mbox{FN}_{\ff{frac}}  + \beta\left| p_L - t_L \right| 
\end{equation}
We did not use the Dice loss ($1 - \mbox{DICE} = (\mbox{FP} + \mbox{FN})/(2\mbox{TP} + \mbox{FP} + \mbox{FN})$) since it provides no training gradient when the box is empty (ie.\ when $\mbox{TP} = 0$). %\begin{equation}
%    1 - \mbox{DICE} = 1 - \frac{2\mbox{TP}}{2\mbox{TP} + \mbox{FP} + \mbox{FN}} = \frac{\mbox{FP} + \mbox{FN}}{2\mbox{TP} + \mbox{FP} + \mbox{FN}} 
%\end{equation}
% raining gradient when no vertebral bone is present  and thus the network can learn to output no segmentation in such cases. 
%Given binary reference labels $t_i$ and probabilistic predictions $p_i$, differentiable expressions for $\mbox{FP}$ and $\mbox{FN}$ are:
%\begin{equation}
%    \mbox{FP} = \sum\limits_{i\in \mathcal{I}} (1 - t_i)p_i  \quad\quad \mbox{FN} = \sum\limits_{i\in \mathcal{I}} t_i(1 - p_i)
%\end{equation} 
The factor $\lambda_n$ in equation \ref{eqn:lossfun} puts a lower weight on false positives during the beginning of training. Since the number of background (empty) voxels is much greater than the number of foreground voxels in the training dataset, at the start of training false positives are a much larger problem than false negatives. If $\lambda_n$ is too large then during training the model ends up just outputting an empty segmentation mask to minimize the loss function. We adjust $\lambda_n$ as a function of the iteration number $n$ starting with $\lambda_n = 0.05$ at $n=0$ and increase it sigmoidally to $\lambda_n \approx 1$ when $n = n_{\ff{max}}$. The factor $\beta$ is a hand-tuned factor which balances the importance of two terms (we settled on $\beta = 1000$). We generated 30\% of our training subvolumes to be empty, and set their ``level'' set to 0. We also tried removing the term $\beta\left| p_L - t_L \right|$ from the loss function (eqn.\ref{eqn:lossfun}) for empty boxes and (independently) adding an additional Dice loss term, but found that neither of these modifications led to an improvement. Training was performed on a single Titan Z GPU with a batch size of 6 and run for $\approx$ 50,000 iterations.

During inference, the network can operate in either ``top-down'' or ``bottom-up'' mode - ie. starting from the top of the image and working down or vice versa. We found top-down to be easier since we can center the network on the spine, which is always present at the top of the image in our dataset. The network starts with a box centered on the average of the bone voxels in the first two slices, which is typically close to the spine in top-down mode. If less than 1 cm$^3$ of vertebral bone is detected in this first subvolume, the algorithm then starts from the top corner of the images and begins scanning across the image with a stride of 64 the x and y directions. When more than a predefined threshold of bone is detected, the subvolume is moved so that it is centered on the detected bone. This process of centering is iterated until the center of the segmented bone changes by less than 2 mm between iterations or until a maximum of 5 iterations is reached. In order to make the algorithm robust, we had to force the algorithm only to move down when operating in top-down mode and only to move up when operating in bottom-up mode. At the end of the centering procedure the segmentation is saved to the memory cell and the regressed level is also saved. Then the next vertebra segmentation is performed, and the centering process is repeated. The algorithm terminates when one of a several different conditions are met. For instance, in top-down mode, the algorithm terminates if it reaches the bottom of the scan, if the predicted level is greater than 24, or if no new bone is found. 

At the end of this process, the segmentation with the level closest to L1 is saved as the predicted level. To output the ordering for the final output segmentation labelmap we perform a max-likelihood calculation to obtain the most likely ordering, where the regression outputs are re-interpreted as likelihoods. A vector of 24 likelihood values is determined for each vertebra. 
%We then define $\delta = L_{\ff{pred}} - L$, where $L_{\ff{pred}}$ is the predicted level and $L$ is $L_{\ff{pred}}$ rounded to the nearest integer. 
For each segmented vertebra $i$ we calculate its likelihood vector $l^i$ using an unnormalized Gaussian function : 
\begin{equation}
    l^i_j = e^{ -\frac{1}{2}((L_{\mbox{\tiny{pred}}} - j)/\sigma)^2 }
\end{equation}
We set $\sigma=2$ (the value used here does not seem to be critical).
%\begin{equation}
%    \begin{aligned}
%        \mbox{If } \delta > 0 \quad &\mbox{:} \quad l_L = 1 - \delta, \quad p_{L+1} = \delta \\
%        \mbox{If } \delta < 0 \quad &\mbox{:} \quad  l_L = 1 + \delta, \quad p_{L-1} = -\delta
%    \end{aligned}
%\end{equation}
Next, all possible orderings consistent with the number of vertebra found are considered. If we store the ordering in a vector $v$ of length $N$, then the likelihood for that ordering is computed as $\mathcal{L}(v) = \prod_i^N l^i_{v_i}$. The ordering with the highest likelihood is chosen as the final ordering. 

%----------------------------------------------------------------------------
\section{Results}
\begin{table}[ht]
\caption{Results from different tests. For the 3D U-Net, The size of the validation sets varied between $N=4$ to $N=10$. $N_{\ff{aug}}$ refers to the number of augmented versions generated for each scan. The average error and median error for the L1 level in the test set are reported in mm. Note this is not a head-to-head comparison as the dataset used for training the 3D U-Net was a different dataset using pseudosegmentations compared to the VerSe data used for training the instance segmentation approach. $^*$ 62 scans were used to generate 8,595 training volumes, including all augmented versions. } 
\label{tab:segtests}
\begin{center}       
\begin{tabular}{c l c c c c c
c} 
method & classes & $N_{\ff{train}}$  & $N_{\ff{aug}}$  & $\mbox{DICE}_{\ff{val}}$ & $\mbox{avg err}_{\ff{test}}$ (mm) &  $\mbox{med err}_{\ff{test}}$ (mm) & $\mbox{L1 acc.}_{\ff{test}}$\\
\hline
\multirow{9}{1cm}{3D U-Net}
& (L1)           & 22   & 256 & 0.60 & 36.9  & 24.5 & 40\%  \\ % 
%&  (L1, T12)      & 22   & 256 &       &    &   &   \%  \\ %  test
& (L1, T12, ribs)& 22   & 256 & 0.62 & 25.0  & 12.5 & 65\%  \\ %test10, 5 in validation
& (L1, T12+ribs) & 22   & 256 & 0.70 & 22.0  & 15.9 & 73\% \\ %test9
& (L1, T12)      & 28   & 256 & 0.50 & 24.4  & 18.1 & 59\%  \\ %test7 4 in validation
& (L1)           & 30   & 256 & 0.80 & 25.8  & 12.0 & 60\%      \\
& (L1)           & 38   & 256 & 0.80 & 16.4 & 9.0  &  65\%    \\ %test3
& (L1)           & 44   & 256 & 0.80 & 14.3 & 5.8  & 70\%  \\ %test4
& (L1)           & 189  & 8   & 0.70 & 8.0  & 3.0  & 88\%  \\ 
& (L1)           & 215  & 8   & 0.88 & 21.0 & 5.7  & 77\%  \\ 
& (L1)           & 249  & 8   & 0.71 & 11.2 & 4.5  & 98\%  \\ 
\hline %& 65.265625 & 35.5
\footnotesize{Instance}& 1 generic class & $62^*$ & n/a & 0.85 & 65.3 & 35.5  & 30\%  \\ 
\end{tabular}
\end{center}
\end{table}

%----------------------------------------------- 
\subsection{Identifying the L1 level}
Results for different tests are shown in table \ref{tab:segtests}. The test set was a set of 40 scans from a completely different dataset of CT colonography (CTC) cases from the University of Wisoconsin Medical Center. When training with 44 hand segmentations the highest accuracy we were able to achieve was 70\%. In roughly 15\% of cases the L1 vertebra would be shifted up to T12 and in 5\% it would be shifted down to L2 (additionally, around 5\% of cases failed, often due to implants or other abnormalities). We then tried doing a 2 class segmentation of L1 and T12 and 2 class segmentations of L1 and T12+rib. However, the multiclass segmentations showed the same failure mode of ``shifting up'' as the single class, so we abandoned that approach.  

We also experimented with a 6 layer FCNN to regress the distance to the L1 level from a subvolume but found that the performance of the network in the validation data during training plateaued with an unacceptably high error of 30 mm. 

\begin{figure}[ht]
    \begin{center}
    \begin{tabular}{c } %% tabular useful for creating an array of images 
    \includegraphics[height=2.2cm]{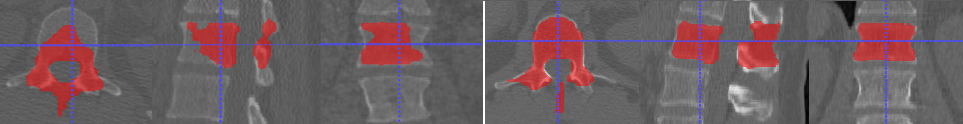} 
    \end{tabular}
    \end{center}
    \caption[] 
    %>>>> use \label inside caption to get Fig. number with \ref{}
    {\label{fig:exampleL1seg} Segmentations of L1 from the U-Net trained on 189 pseudo-segmentations, showing the axial, sagittal, and coronal views. The left image is a CTC case, while the right one is a hold-out case from the Naval data, the dataset used for training.}
\end{figure} 

%---------------------------------------------- 
\subsection{Iterative instance segmentation}
\begin{figure}[ht]
    \begin{center}
       % \begin{tabular}{c} %% tabular useful for creating an array of images 
            \includegraphics[height=4.5cm]{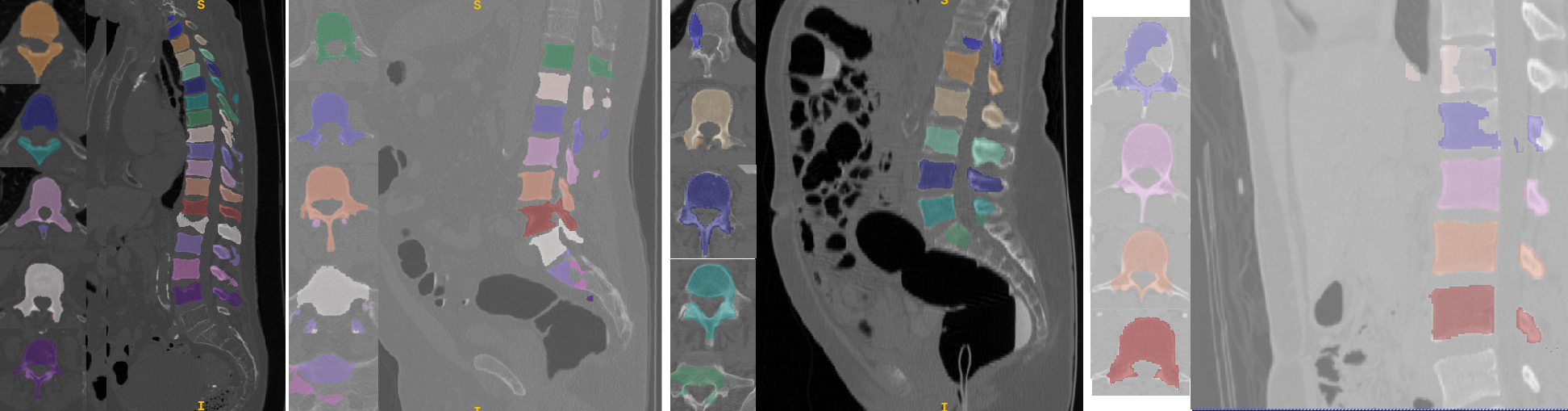}  
      %  \end{tabular}
    \end{center}
    \caption[] 
    %>>>> use \label inside caption to get Fig. number with \ref{}
    {\label{fig:examplesegs} Segmentations of the spine produced using the iterative-instance based method randomly chosen from four datasets. From left to right:  VerSe sample (in-sample ground truth dataset), CT Colonography case from the University of Wisconsin Medical Center, CT Colonography case from National Navy Medical Center,\cite{Pickhardt2003CTCData} and a renal donor case in the post-contrast phase. While the algorithm shows good generalization performance to the different dataset and is capable of working on scans with arbitrary field of view, several weaknesses are apparent such as trouble segmenting the topmost and bottomost vertebrae.}
\end{figure} 

We first tried to train the iterative instance based approach on the same type of pseudosegmentations we used to train the L1 segmentation U-Net. The algorithm requires that the output threshold be tuned afterwards for optimum performance. Several segmentations, which are randomly drawn (ie.\ not hand picked) are shown in figure \ref{fig:examplesegs}. One issue is apparently is that the method often has trouble segmenting the first vertebra because the instance memory is empty and cases with an empty instance memory are rare in the training dataset. One possible trick to overcome this problem would be to to train two separate models for both bottom-up and one for top-down inference, and then combine the bottom-up and top-down segmentation outputs. Another solution would be to create additional training data with the vertebra to be segmented at the top or bottom of the subvolume and an empty instance memory. Quantitative results on the test set of 40 CTC cases are presented in table \ref{tab:segtests}. While this method is not as accurate at identifying L1, qualitatively the segmentations in the CTC and other test datasets are of higher quality than with the direct approach. This is not surprising given that the instance based U-Net was exposed to many more examples of vertebra during training. 
%After observing issues with the first segmentation, we also modified the training procedure so that in 10\% of training examples the memory mask would be empty. This increased the number of training subvolumes where the memory mask was empty, forcing the model to become better at doing the first segmentation. 
%\begin{table}[ht]
%    \caption{Comparison of the L1 level detection accuracy for the iterative instance based approach %(full spine seg) and the direct segmentation approach. Both were trained on pseudo-segmentations from the same dataset of 216 Naval CT colonography cases.} 
%\label{tab:fullspinedice}
%    \begin{center}       
%        \begin{tabular}{l c c} 
%method  & avg err$_{\ff{test}}$ (mm) & L1 acc$_{\ff{test}}$  \\
%\hline
%L1 seg          &           & & \\
%full spine seg &       & & \\
%    \end{tabular}
%  \end{center}
%\end{table}
%\begin{table}[ht]
%    \caption{} 
%    \label{tab:fullspinedice}
%    \begin{center}       
%        \begin{tabular}{l c c} 
%name of test &  \% empty & \\
%\hline
%test_               & 25\%   % segs smearing over vertebrae
%test_class_verse_aug & 18\% \\  %#train, #val 3404 17
%    \end{tabular}
%  \end{center}
%\end{table}

%-------------------------------------------------------------- 
\section{Conclusion}
We have compared two very different approaches to identifying the L1 level. First we trained a 3D U-Net to segment the L1 vertebra directly. As far as we know this is the first time this approach has been studied. By training with 249 pseudosegmentations we obtained a model which achieved an accuracy of 98\% at identifying L1 in an independent test dataset of 40 scans. We also developed an iterative instance based segmentation system to segment and classify the entire spine. We hypothesized this approach may be useful when there are vertebral anomalies present in the scan such as additional or missing vertebra. We first trained using the same pseudosegmentations used for training the U-Net for direct segmentation, but we found the results to be of poor quality. Next we trained with hand drawn segmentations from the VerSe dataset. The resulting model yielded more accurate segmentations of nearly the entire spine but a less accurate classification accuracy for L1. While our main goal here was to identify L1, the high quality segmentations obtained by the iterative instance based approach may be useful for performing a measurement of bone mineral density, for instance by performing an erosion on the segmentation and averaging the Hounsfield units within the trabecular space of the L1 vertebra, similar to Liu et al.\cite{Liu2018SPIE} 

%---------------------------------------------------------------------------------- 
\acknowledgments % equivalent to \section*{ACKNOWLEDGMENTS}       
We thank Dr.\ Perry J.\ Pickhardt and Dr.\ J.\ Richard Choi for providing the CTC images used for testing. This research was supported in part by the Intramural Research Program of the National Institutes of Health Clinical Center. 

%-------------------------------------References -------------------------------------------- 
\bibliographystyle{spiebib} % makes bibtex use spiebib.bst
\bibliography{bibliography.bib}  

\end{document}